# Improving hot-spot pressure for ignition in high-adiabat inertial confinement fusion implosion

Dongguo Kang(康洞国),* Shaoping Zhu(朱少平),† Wenbing Pei(裴文兵), Shiyang Zou(邹士阳), Wudi Zheng(郑无敌), Jianfa Gu(谷建法), and Zhensheng Dai(戴振生)

*Institute of Applied Physics and Computational Mathematics*

*Beijing 100088, China*

(Dated: February, 2017)

A novel capsule target design to improve the hot-spot pressure in the high-adiabat implosion for inertial confinement fusion is proposed, where a layer of comparatively high-density material is used as a pusher between the fuel and the ablator. This design is based on our theoretical finding of the stagnation scaling laws, which indicate that the hot spot pressure can be improved by increasing the kinetic energy density $\rho_d V_{imp}^2/2$ ($\rho_d$ is the shell density when the maximum shell velocity is reached, $V_{imp}$ is the implosion velocity.) of the shell. The proposed design uses the high density pusher to enhance the shell density $\rho_d$ so that the hot spot pressure is improved. Radio-hydrodynamic simulations show that the hot spot pressure of the design reaches the requirement for ignition even driven by a very high-adiabat short-duration pulse. The design is hopeful to simultaneously overcome the two major obstacles to achieving ignition——ablative instability and laser-plasma instability.

For the ignition of inertial confinement fusion (ICF) [1–3], it is cardinal to assemble a hot spot with high pressure. The energy required for ignition $E_{ign}$ approximately scales with the hot-spot pressure $P_{hs,\tau}$ as $E_{ign} \sim P_{hs,\tau}^{-2}$ [1]. To achieve ignition with affordable laser energy, the capsule is usually designed to obtain high $P_{hs,\tau}$ by both accelerating the deuterium-tritium (DT) fuel shell to high velocity and keeping it on a low adiabat [4] by a four-step low-adiabat driving pulse. But the low-adiabat implosion is sensitive to the ablative hydrodynamic instability so that the experiments of the low-adiabat design on the National Ignition Facility (NIF) suffered from severe hot-spot mix which prevents the achievement of ignition [5–7]. By imploding the ablator on a higher adiabat, the ablative instability is significantly mitigated [8, 9]. So, the experiments on NIF acquire low hot-spot mix and achieved the key step of fuel gain exceed unity by using the design driven with a three-step high-adiabat pulse [11]. While, the high-adiabat pulse also sets the fuel shell on a higher adiabat and results in reduced $P_{hs,\tau}$ and higher $E_{ign}$, Therefore ignition can not be achieved on NIF yet through the high-adiabat implosion using nowadays target design. In this paper, we propose a novel target design scheme to improve the $P_{hs,\tau}$ in the high-adiabat implosion. This scheme replaces a part of the fuel shell with a layer of comparative higher density material and places the high density layer between the remainder fuel and the ablator as a pusher. Numerical simulations show that the design by this scheme improves the $P_{hs,\tau}$ by a factor near 70% comparing to the usual target design and achieves ignition in a very high-adiabat implosion driven by a two-step pulse. This scheme is based on the theoretical finding that the hot-spot pressure is mainly determined by the kinetic energy density $\rho_d V_{imp}^2/2$ of the shell where $V_{imp}$ is the maximum shell velocity (implosion velocity) and $\rho_d$ is the shell density at its maximum velocity. In the high-adiabat implosion, the $P_{hs,\tau}$ is reduced due to the decreased $\rho_d$ because of the shell's higher entropy, while the new scheme enhances $\rho_d$ by utilizing the high density pusher. For convenience, we start our discussion from the hydrodynamic laws of the implosion.

In the implosion, the shell of a capsule is accelerated to high velocity then compresses the central hot-spot in the deceleration phase. The maximum hot-spot pressure is reached at the end of the deceleration phase when the imploding shell stagnates. The process before deceleration can be directly controlled through the capsule characters and the driving source specifications. After the deceleration begins, the pressure in the capsule is much higher than the driving pressure, so the implosion could almost not longer be affected by the external conditions and is governed by the intrinsic hydrodynamic laws. Thereby, it is critical to obtain the intrinsic hydrodynamic laws and the relations between the stagnation parameters and the in-flight quantities (quantities at the beginning of the deceleration phase) for understanding the physical factors that determines the compression.

At stagnation, the system consists of a hot spot surrounded by the "cold" shell. Each of them is fully described by three parameters, two describing the state, such as temperature and density, one describing the dimension, such as radius. According to the isobaric model [1, 12], the pressures of the two parts are approximately equal, then five parameters are sufficient for determining the assembly. Investigations had been made on the hydrodynamic laws of the deceleration phase. Kemp *et al.*

[13] derived two scaling laws describing the stagnation pressure and shell density basing on Guderley's self-similar flow theory. Betti *et al*. [14] developed a "thin shell model"(TNSM) and a "thick-shell model"(TKSM) of the deceleration phase. TKSM is closer to the reality but more complex. Combining the TNSM model with the isobaric model and the self-similar flow hot-spot model [16], Zhou *et al*. [15] derived scaling laws with five stagnation parameters. But neither Kemp's nor Zhou's scaling laws can be completely in quantitative agreement with numerical simulations. Here we base on the TKSM to derive the scaling laws for the stagnation parameters.

The TKSM simplified the dynamics of the deceleration phase as "one-shock" process which supposes that only one shock returns from the center and propagates out through the shell. The return shock divides the shell into two parts as shown in the Fig. 2 of Ref [14]: the shocked shell which is inside the return shock and the free-fall shell which is outside. The shocked shell acts like a piston on the hot spot and the free-fall shell provides compression work rate through the flow of momentum across the shock. The compression ends when the shocked shell exhausts its imploding kinetic energy and stagnates. This model is described by the set of ordinary differential equations (22), (23), (25) and (26) in Ref [14]. This model considers the actual free-fall shell density profile $\rho_{ff}(r)$ which is described by Equ. (28) in Ref [14]. This enormously complicates the acquirement of analytical solutions. But we find that the effect of the actual density profile on the compression is approximately equivalent to a uniform density profile. The free-fall shell acts on the compression through providing compression work. As the velocity in the free-fall shell is nearly uniform, the provided compression work is proportional to the mass of the shocked part $M_{ss} \approx 4\pi R_0^2 (\int \rho_{ff} dr)_{shocked}$ and is determined by $(\int \rho_{ff} dr)_{shocked}$. As shown in Fig. 1, the $\int \rho_{ff} dr$ of the actual density profile is well approximate to a uniform density profile when $M_{ss}/M_{sh}$ is in the range of 0.2~0.85, here $M_{sh} \approx 4\pi R_0^2 (\int \rho_{ff} dr)_{\Delta_0}$ is the total mass of the shell.

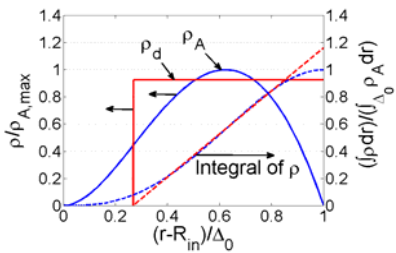

Fig. 1. Comparison of the integral $\int \rho dr$ between an actual shell density profile and an equivalent uniform density profile. $\rho_A$ and $\rho_d$ denote the actual and the equivalent density profile. The $R_{in}$ and $\Delta_0$ in the X-axis label is the inner radius and the thickness of the actual shell. The density $\rho$ is normalized by the maximum value of the actual density $\rho_{A,max}$. The integral $\int \rho dr$ is normalized by the integral of $\rho_A$ over the whole shell.

This suggests that the effect of the actual free-fall shell on the compression is approximately equivalent to a uniform density shell when $0.2 \leq M_{ss,\tau}/M_{sh} \leq 0.85$ where $M_{ss,\tau}$ is the value of $M_{ss}$ at stagnation. Actually, $M_{ss,\tau}$ is usually higher than $0.2M_{sh}$ in the implosion. Therefore, the approximation of a uniform density shell requires $M_{ss,\tau} \leq 0.85 M_{sh}$.

Using the equivalent uniform free-fall density (denoted with $\rho_d$), the equations of the TKSM can be written as the following dimensionless form without considering the alpha-particle energy deposition:

$$\frac{d\hat{P}_{hs}}{d\hat{t}} + 3\gamma \frac{\hat{P}_{hs}}{\hat{R}_{hs}} \frac{d\hat{R}_{hs}}{d\hat{t}} = 0 , \qquad (1)$$

$$\frac{d\hat{M}_{ss}}{d\hat{t}} = \left(\frac{d\hat{R}_k}{d\hat{t}} + 1\right)\hat{R}_k^2 , \qquad (2)$$

$$2\hat{\xi}_0 \frac{d}{d\hat{t}}\left[\hat{M}_{ss}\left(\frac{\hat{R}_{hs}+\hat{R}_k}{2\hat{R}_{hs}}\frac{d\hat{R}_{hs}}{d\hat{t}}+1\right)\right] = \hat{P}_{hs}\hat{R}_{hs}^2 , \qquad (3)$$

$$\frac{\gamma+1}{2}\left(\frac{\hat{R}_k}{\hat{R}_{hs}}\frac{d\hat{R}_{hs}}{d\hat{t}}+1\right) = \frac{d\hat{R}_k}{d\hat{t}}+1 , \qquad (4)$$

where γ is adiabatic index, the dimensionless hot-spot pressure $\hat{P}_{hs}$ and radius $\hat{R}_{hs}$, position of return shock $\hat{R}_k$ and shocked-shell mass $\hat{M}_{ss}$ are defined as

$$\hat{P}_{hs}=\frac{P_{hs}}{P_0}, \quad \hat{R}_{hs}=\frac{R_{hs}}{R_0}, \quad \hat{R}_k=\frac{R_k}{R_0}, \quad \hat{M}_{ss}=\frac{M_{ss}}{4\pi\rho_d R_0^3}. \qquad (5)$$

The dimensionless time $\hat{t}$ is defined as

$$\hat{t}=tV_{imp}/R_0 , \qquad (6)$$

and $\hat{\xi}_0$ is a dimensionless parameter defined as,

$$\hat{\xi}_0=\frac{\rho_d V_{imp}^2/2}{P_0} . \qquad (7)$$

In (5)~(7), $P_0$ and $R_0$ are the hot-spot pressure and radius at the beginning of deceleration phase, $V_{imp}$ is the implosion velocity. Using the initial conditions:

$$\hat{P}_{hs}(0)=1, \quad \hat{R}_{hs}(0)=1, \quad \hat{R}_k(0)=1, $$
$$\frac{d\hat{R}_{hs}}{d\hat{t}}(0)=-1, \quad \hat{M}_{ss}(0)=0 , \qquad (8)$$

the following stagnation solutions are derived:

$$\hat{P}_{hs,\tau}=\Phi\hat{\xi}_0^{\frac{\gamma}{2(\gamma-1)}} , \qquad (9)$$

$$\hat{R}_{hs,\tau}=\Theta\hat{\xi}_0^{\frac{-1}{6(\gamma-1)}} , \qquad (10)$$

$$\hat{R}_{k,\tau}=\Omega\hat{\xi}_0^{\frac{-1}{6(\gamma-1)}} , \qquad (11)$$

$$\hat{M}_{ss,\tau}=\Psi\hat{\xi}_0^{\frac{1}{2}} . \qquad (12)$$

Here the subscript "τ" denotes the stagnation quantities.

The coefficients Φ, Θ, Ω and Ψ in (9)~(12) are constants determined by γ. In deriving the solutions, the conditions γ< 2 and $\hat{\xi}_0 \gg 1$ are required, which is suitable for ICF implosions. As described previously, five independent parameters are required to fully describe the stagnation assembly. Here only four are derived. The stagnation hot-spot density $\rho_{hs,\tau}$ is selected as an additional parameter. According to the dimensional relation of $P_{hs,\tau}$, $\rho_{hs,\tau}$ and $V_{imp}$ which requires $P_{hs,\tau}/\rho_{hs,\tau} \sim V_{imp}^2$, we obtained the scaling law for $\rho_{hs,\tau}$ as

$$\hat{\rho}_{hs,\tau} = \Gamma \hat{\xi}_0^{\frac{2-\gamma}{2(\gamma-1)}}, \quad (13)$$

where Γ is a constant depending on γ.

It is interesting to notice that all the dimensionless stagnation quantities given by Eqs.(9)~(13) are uniquely determined by $\hat{\xi}_0$. This is understandable as the numerator $\rho_d V_{imp}^2/2$ of $\hat{\xi}_0$ is the shell's kinetic energy density which represents the shell's ability to compress the hot spot and the denominator $P_0$ represents the compressibility of the hot spot.

Numerical simulations are carried out to verify the scaling laws. The simulations use a simplified capsule model which is similar to that used by Herrmann to numerically generate the scaling law of ignition energy in Ref. [17]. The model considers the fuel part of the real capsule. It consists of a spherical DT shell filled with low density DT gas (0.3mg/cm³). Implosions are driven by a pressure pulse acting on the outer surface of the shell. The top pressure of the driving pulse is denoted with $P_d$. Simulations are carried out using a one-dimensional radiation-hydrodynamics code without considering alpha-heating. The implosions with $M_{ss,\tau}/M_{sh} \geq 0.85$ are selected according to the requirement of the uniform density approximation. The simulation results are plotted in Fig. 2 and compare to the scaling laws in Equ. (9)~(13). As shown, the analytical scaling laws are in good agreement with the simulation results as a whole.

As described previously, the uniform density approximation holds with $M_{ss,\tau}/M_{sh} \leq 0.85$. When $M_{ss,\tau} > 0.85 M_{sh}$, the much lower density in the actual free-fall shell will result in lower $P_{hs,\tau}$. If the whole shell is shocked by the return shock before stagnation, a rarefaction will generate after the return shock reaches the outer surface of the shell. This rarefaction will decompress the assembly and further lower down the $P_{hs,\tau}$. All these results are verified by the simulation results shown in Fig. 3. This time, a set of implosions with approximately the same in $\rho_d$ and $P_0$ and difference in $V_{imp}$ is generated. For these implosions, equ. (9) results in $P_{hs,\tau} \propto V_{imp}^2$. The dependence of $P_{hs,\tau}$ and $M_{ss,\tau}/M_{sh}$ on $V_{imp}$ is shown in Fig. 3a and 3b respectively. As predicted, the simulated $P_{hs,\tau}$ deviates from the scaling law and tends lower when $M_{ss,\tau}/M_{sh} > 0.85$. Another conclusion can be deduced from the analytical model and the simulations: only a part of the shell takes effective to the compression if the shell is sufficient thick. The mass of the effective part is described by $M_{ss,\tau}$.

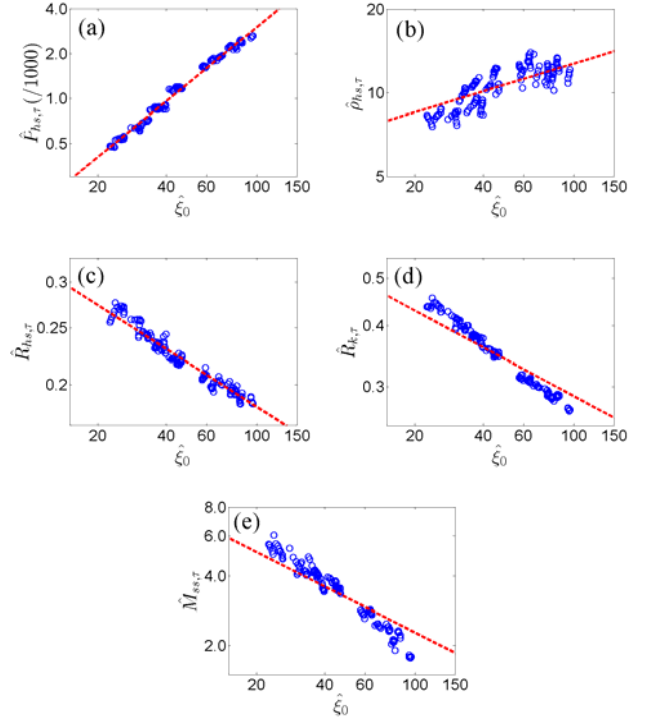

FIG. 2. The comparision of the scaling laws given by Eqs.(9)~(13) with the simulation resutls where γ=5/3.

As described previously, to achieve ignition is to achieve sufficient high $P_{hs,\tau}$. Substituting Eq. (7) into Eq. (9) and taking γ=5/3, one have

$$\hat{P}_{hs,\tau} = \Phi \left( \frac{\rho_d V_{imp}^2}{2} \right)^{5/4} P_d^{-1/4} \quad (14)$$

where $P_0 \approx P_d$ is used as deceleration begins when the hot-spot pressure exceeds the driving pressure. It indicates that $P_{hs,\tau}$ is mainly improved by increasing $\rho_d V_{imp}^2/2$, as the scaling power of $P_d$ is small and $P_d$ is mainly determined by driving radiation temperature which varies in a relatively small range (usually 250eV~350eV). According to our numerical simulations of CH capsule design, it is found that $\rho_d V_{imp}^2/2 \sim 1.6$(GJ/cm³) for the conventional low-adiabat implosion with $\alpha_{if} \sim 1.5$ and $V_{imp} \sim 370$km/s, but $\rho_d V_{imp}^2/2 \sim 0.83$(GJ/cm³) for the high-foot implosion with $\alpha_{if} \sim 2.5$ and $V_{imp} \sim 380$km/s, here the adiabatic factor $\alpha_{if}$ is defined as the ratio of the pressure to the Fermi-degenerated pressure at the beginning of the deceleration phase. It is obviously that there are two ways to enhance $P_{hs,\tau}$, one is to increase density $\rho_d$ and the other is to increase velocity $V_{imp}$. Increasing $V_{imp}$ usually encounters higher in-flight aspect ratio and is limited by the ablative instability growth [2]. There are several methods to increase $\rho_d$, one of which is the low-adiabat implosion. But as mentioned above, the ablative instability growth for the low-adibat implosion is intolerably high. It is very helpful to propose a method to increase $\rho_d$ for the high-adiabat implosion.

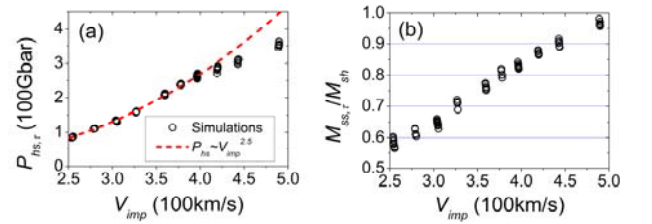

FIG. 3. The impact of the ratio of the shocked-shell mass to the total shell mass $M_{ss,\tau}/M_{sh}$ on the scaling law of the stagnation hot-spot pressure $P_{hs,\tau}$. a) the dependence $P_{hs,\tau}$ on $V_{imp}$, b) the corresponding $M_{ss,\tau}/M_{sh}$ of the implosions with different $V_{imp}$.

In the implosion of a usual capsule, the shell effective to compression in the deceleration phase mainly consists of DT fuel. As the shell comes from DT ice whose initial density (0.255g/cm$^3$) is low, it is imploded on a low adiabat to obtain enough high $\rho_d$. But if the initial density of the shell is high enough, high $\rho_d$ can also be obtained even driven by a high-adiabat pulse. For example, as the shell is isoentropically accelerated, the $\rho_d$ can be obtained as $\rho_d = (P_d/P_a)^{1/\gamma}\rho_a$ where $P_a$ and $\rho_a$ are the shell pressure and density at the beginning of acceleration phase. $P_d/P_a$ is correlated with spherical convergent effect and $\rho_d$ is primarily determined by $\rho_a$. A low-adiabat four-shock pulse sets $\rho_a$ to ~8g/cm$^3$ according to the relation $P_a = 2.17\alpha_{if}\rho_a^{3/5}$ for typical $P_a$ ~100Mbar and $\alpha_{if}$ ~1.5, while $\rho_a$ can be set to 8g/cm$^3$ even driven by an one-shock pulse if the initial density of the shell is 2.0g/cm$^3$ (taking γ=3/5). Such high initial density is of course impossible for DT. So we propose a layer of comparative high initial density material to replace a part of the DT fuel and be placed between the remainder DT ice and the ablator as a pusher. Both the DT fuel and the pusher compose the shell effective to compression. As the pusher raises the overall initial density of the shell, higher $\rho_d$ is acquired, and the $P_{hs,\tau}$ is improved.

Noticing that $P_{hs,\tau}$ is also correlated with the properties of the shell material which is represented by γ in equation (9), proper material should be selected as the pusher. Another important issue about the pusher material is that its albedo should be high enough. Otherwise, its density may be severely lowered down due to the preheat of the high frequency X-ray during the acceleration phase. Several materials with density near 3.0g/cm$^3$ have been tested for the pusher by simulations, and aluminum (Al, 2.7g/cm$^3$) is found to be a suitable pusher material.

A capsule design using Al as the pusher material and the driving radiation pulse are shown in Fig. 4. The outer radius of the capsule is 1072μm. It uses a layer of 7μm aluminum as the pusher and high density carbon (HDC) [18] as the ablator. Because of the existence of the pusher, a layer of DT ice with only 30μm is filled to reserve enough high implosion velocity and ensure that the pusher is effective to the compression. Implosion is driven by a two-step radiation pulse with foot temperature 145eV and top temperature 300eV. Simulations of implosion are performed. The performance is listed in Tab. I and compared to the simulated results of a NIF-like HDC capsule [19] driven by the same two-step pulse (the main pulse is optimally timed). Although the pusher capsule implode with a lower $V_{imp}$, its $\rho_d V_{imp}^2/2$ is beyond 50% larger than the NIF like capsule, which results in ~69% higher $P_{hs,\tau}$. Ignition is achieved in the implosion of the pusher target with 8.20MJ yield which is relative low due to small fuel mass (0.09mg). Whereas, the NIF like capsule only generates 1.87MJ fusion yield. The simulations clearly show significant effect of the Al pusher to the improvement of the hot-spot pressure and so the implosion performance.

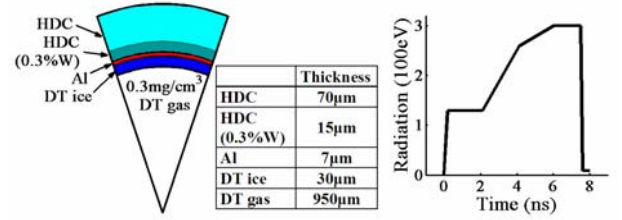

FIG. 4. The capsule and driving radiation pulse for pusher target design

The advantage of the pusher target design is that it can use a very high-adiabat driving pulse to mitigate the ablative instability. The other potential advantage lies in that the driving laser pulse is very short. This is favorable for using a near-vacuum hohlraum (NVH) to drive the implosion, where the laser-plasmas interaction is greatly mitigated. The experiments [20] of NVH on NIF showed ~40% higher hohlraum efficiency and better symmetry control than the gas-filled hohlraum which is used for the usual ignition target with a long driving pulse. As the density of the filled gas in the NVH is very low, the plasmas ablated by laser from the hohlraum wall expand in a much faster speed. If the laser pulse is too long, the ablated plasmas would expand into the laser path and hinder the laser propagation. Experiments on NIF have proved high energy coupling efficiency in the NVH driven by a laser pulse with time duration up to 8ns [21]. Noticing that the driving pulse of the pusher target design is only 7.5ns, a NVH can be used to driven the implosion in principle.

The primary risk of the pusher target design comes from the hydrodynamic instability of the pusher's surfaces. Simulation indicates that the outer surface of the Al pusher is Rayleigh-Taylor(RT) [22, 23] unstable before decelerating and stable after decelerating because its density exceeds the ablator from later acceleration. Before decelerating, the inner surface of the Al pusher is RT stable as its density is higher than the fuel. After decelerating, the inner surface is surprisingly found to be RT stable too. The reason lies in that the fuel is shocked to higher density than the pusher by the return shock which begins to decelerate the pusher and the higher density of the fuel is kept to the end of the deceleration. As the result, only the outer surface of the Al pusher in the acceleration phase is RT unstable. Two-dimensional simulations with perturbations initially set on the interface of the Al/HDC were performed to assess the instability growth. The growth factor (GF) of the perturbations with mode number high up to 512 were obtained as shown in Fig. 5. The highest GF is predicted to ~200 from the simulations. The instability is at a low level comparing to that of the interface between DT fuel and the ablator in the low-adiabat implosion of the usual capsule where the highest GF is predicted to ~1500 [9].

TABLE I. Implosion performance of the pusher capsule and the NIF like capsule driving with the same two-step radiation pulse shown in Fig.4.

| Capsule type | Pusher capsule | NIF-like capsule |
|---|---|---|
| $V_{imp}$ (km/s) | 367[a] | 403 |
| $\alpha_{if}$ | 2.86 | 3.05 |
| $\rho_d V_{imp}^2/2$ (GJ/cm$^3$) | 1.52[a] | 1.00 |
| $P_{hs,\tau}$ (Gbar) | 471 | 279 |
| Yield (MJ) | 8.20 | 1.87 |
| Burn fraction | 26.7% | 1.35% |

[a] Averaging over the fuel shell and the pusher

In summary, we have theoretically obtained the stagnation scaling laws for the compression of the deceleration phase and found that the hot-spot pressure are primarily improved by increasing the shell's kinetic energy density at peak velocity $\rho_d V_{imp}^2/2$. We also proposed a novel target design which uses a comparative high density pusher to enhance $\rho_d$ and so that improve the hot-spot pressure to the requirement for ignition in the very high-adiabat implosion. It is a promising design scheme as it is robust to both ablative instability and laser plasmas instabilities which are two major obstacles to achieving ignition.

This work is supported by the National Natural Science Foundation of China under Grant No. 11375032, the Science and Technology Foundation of CAEP under Grant No. 2015B0202035.

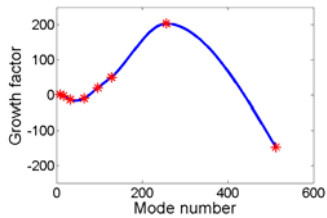

FIG. 5. Instability growth factors of the outer surface of the Al pusher at peak velocity. The growth factors were obtained by two-dimensional simulations of the pusher target design in Fig. 4 with perturbations initially set on the outer surface of the Al pusher.

*kang_dongguo@iapcm.ac.cn
†zhu_shaoping@iapcm.ac.cn